\documentclass{osa-article}

\journal{oe}
\articletype{Research Article}
\usepackage{extarrows}
\usepackage{mathrsfs,amsmath}

\begin{document}

\title{X-ray phase-sensitive microscope imaging with a grating interferometer: theory and simulation}

\author{{\color{black}Jiecheng Yang,\authormark{1} Yongshuai Ge,\authormark{1,2,*} Dong Liang,\authormark{1,2} and Hairong Zheng\authormark{2}}}

\address{\authormark{1}Research Center for Medical Artificial Intelligence, Shenzhen Institute of Advanced Technology, Chinese Academy of Sciences, Shenzhen, Guangdong 518055, China\\
\authormark{2}Paul C Lauterbur Research Center for Biomedical Imaging, Shenzhen Institute of Advanced Technology, Chinese Academy of Sciences, Shenzhen, Guangdong 518055, China}

\email{\authormark{*}ys.ge@siat.ac.cn} 


\begin{abstract*}
In this work, a general theoretical framework is presented to explain the formation of the phase signal in an X-ray microscope integrated with a grating interferometer, which simultaneously enables the high spatial resolution imaging and the improved image contrast. Using this theory, several key parameters of phase contrast imaging can be predicted, for instance, the fringe visibility and period, the conversion condition from the differential phase imaging (DPI) to the phase difference imaging (PDI). Additionally, numerical simulations are performed with certain X-ray optical components and imaging geometry. Results demonstrate the accuracy of this developed quantitative analysis method of X-ray phase-sensitive microscope imaging.

\bigskip

\end{abstract*}

%%%%%%%%%%%%%%%%%%%%%%%%%%  body  %%%%%%%%%%%%%%%%%%%%%%%%%%
\section{Introduction}
The development of phase-sensitive X-ray imaging techniques over the last decades allows the measurement of the inner structure of weakly absorbing objects (e.g., soft tissue, carbon materials) with high sensitivity, and dramatically complements the conventional absorption imaging. Among the various promising X-ray phase contrast imaging methods \cite{Momo03, Pfei06, Zhu10, Zhu14, Momo20, Xu20, Zan21}, the grating-based X-ray Talbot and Talbot-Lau interferometers have been subject to increasing attention due to its compatibility with X-ray tube imaging systems. Many efforts \cite{Take08, Yash09, Yash10, Kuwa11, Beru12, Taka17} have been taken by integrating the Talbot(-Lau) interferometry with the full-field transmission X-ray microscope. One exciting progress is demonstrated by Takano et al. \cite{Taka17, Taka18} on such a system to obtain superior phase information in comparison with the Zernike phase-contrast imaging approach. However, instead of generating the DPI images, such a combined X-ray microscope system produces the PDI images, which need to be post-processed via the iterative deconvolution method \cite{Taka19} or the maximum likelihood reconstruction method \cite{Wolf20} to recover the phase information. Yashiro et al. \cite{Yash09, Yash10} have provided pioneering theoretical explanations for such PDI phenomenon by exploiting the Talbot self-imaging effect.

In this work, a more general theoretical analysis for the entire imaging procedure that starts from the source and ends on the detector is proposed. The new theory is able to deal with any shaped source, and thus permits the developments of innovative interferometer designs. Additionally, the conversion condition from the PDI to the DPI, or vise versa, is  quantitatively investigated with respect to the resolution limit of the imaging system. Finally, numerical simulations are performed to verify the consistency between the theoretical predictions and the previous experimental observations \cite{Taka18}.

\section{Theoretical framework}
\label{sec:theory}

\begin{figure}[t]
\centering\includegraphics[width=0.82\linewidth]{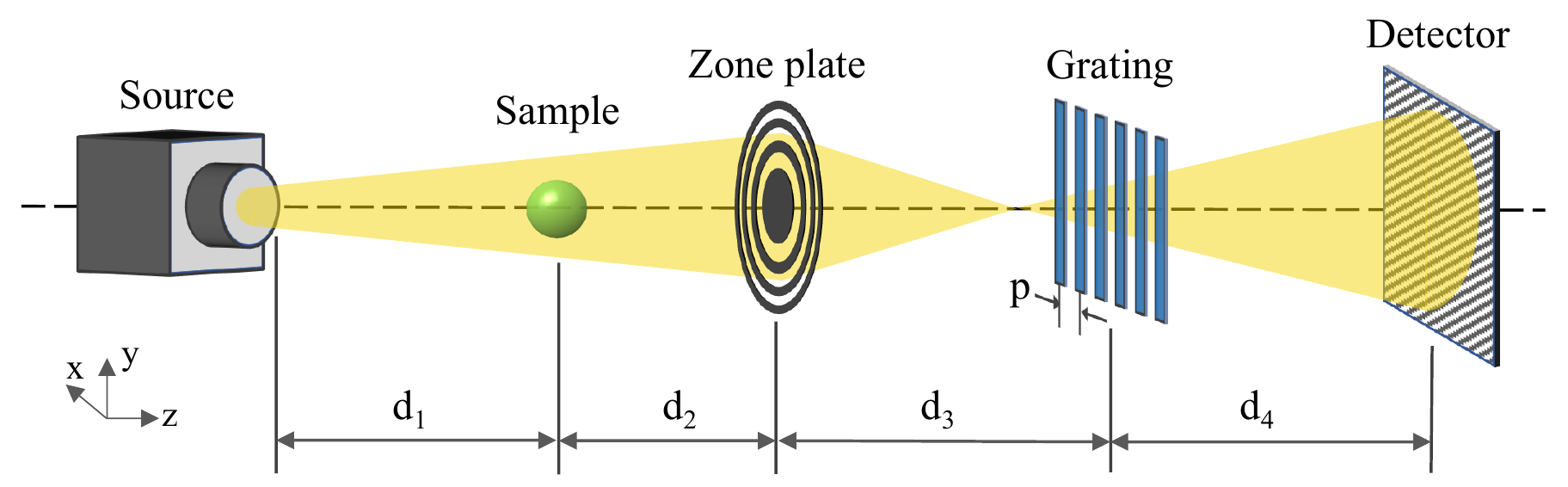}
\caption{Illustration of the X-ray phase contrast microscope imaging system with a point source and a $\frac{\pi}{2}$ phase grating.}
\label{fig:fig1}
\end{figure}

The assumed X-ray microscope imaging system is depicted in Fig.~\ref{fig:fig1}. Herein, a spatially coherent and quasi-monochromatic X-ray point source is assumed. The object distance $d_2$ and the image distance $\left(d_3+d_4\right)$ satisfy the thin lens equation: $\frac{1}{d_2}+\frac{1}{d_3+d_4}=\frac{1}{f}$, in which $f$ represents the focal length of the focusing device, e.g., the zone plate. The value of $M=\left(d_3+d_4\right)/d_2$ defines the magnification ratio of the system. In order to acquire the phase information besides the absorption signals, a $\frac{\pi}{2}$ phase grating is added between the zone plate and the detector. 

In this study, the X-ray propagation is governed by Fresnel diffraction, which is a near-field approximation of the Kirchhoff-Fresnel diffraction for scalar waves. To simplify the derivations, the following analyses will be conducted in a one-dimensional (1D) case (along the $x$ axis). The two-dimensional (2D) results can be easily obtained upon this basis. With the paraxial approximation, the diffracted field $U_{\text{out}}(x')$ at any distance $d$ from the initial wave field $U_{\text{in}}(x)$ can be expressed as:
\begin{equation}
\label{eq:eq1}
\begin{aligned}
U_{\text{out}}(x') & =\frac{1}{\sqrt{i\lambda d}}\int U_{\text{in}}(x) e^{ik\left(d+\frac{(x'-x)^2}{2d}\right)} dx \\
& = \sqrt{\frac{2\pi}{i\lambda d}}\mathscr{F}^{-1}\left(\mathscr{F}\left(U_{\text{in}}(x)\right)\mathscr{F}\left(e^{ik\left(d+\frac{x^2}{2d}\right)}\right)\right), 
\end{aligned}
\end{equation}
where $\lambda$ is the X-ray wavelength, and $k$ (=$\frac{2 \pi}{\lambda}$) is the wave number. $\mathscr{F}$ and $\mathscr{F}^{-1}$ denote the Fourier transform and inverse Fourier transform, respectively. The Eq.~(\ref{eq:eq1}) plays a fundamental role in obtaining the beam intensity at different positions, and would be used repeatedly during the following derivations.

Using Eq.~(\ref{eq:eq1}), the scalar wave field $U_1(x_1)$ of a point source $\delta(x_0-\eta)$ ($\eta$ denotes the off-axis distance) after propagating freely over a distance $d_1$ (before the sample) can be calculated as 
\begin{equation}
\label{eq:eq2}
U_1(x_1) = \frac{1}{\sqrt{i \lambda d_1}} e^{i k  \left(d_1+\frac{(x_1-\eta)^2}{2 d_1}\right)}.
\end{equation}
Assuming that the thickness of the sample along the optical axis is negligible compared to the focal length of the zone plate, the modulation of the X-ray wavefront after penetrating the sample with a refractive index $n=1- \delta + i\beta$ is approximated by  
\begin{equation}
\label{eq:eq3}
U'_1(x_1) =U_1(x_1)e^{- k \alpha(x_1)-i k \phi(x_1)}, 
\end{equation}
where $\alpha(x_1)=\int\beta(x_1, z_1) dz_1$ and $\phi(x_1)=\int\delta(x_1, z_1)dz_1$ represent the decay and phase shift of the wave amplitude, respectively. Considering a pure phase object, i.e., $\alpha(x_1)=0$, its phase shift $\phi(x_1)$ is further supposed to be smooth, continuous and slowly varying. Hence, its derivative, essentially the angle of refraction, always exists. Based on these assumptions, the amplitude of the wave field in front of the zone plate becomes 
\begin{equation}
\label{eq:eq4}
U_2(x_2)  \approx \frac{  e^{i k \left( d_1 + d_2 + \frac{ (x_2-\eta)^2}{2 \left(d_1+d_2\right)}- \phi \left(\frac{\eta d_2+d_1 x_2}{d_1+d_2}\right)\right) } }{\sqrt{i \lambda \left(d_1+d_2\right) }}. 
\end{equation}
With the projection approximation, the wavefront after being modulated by the zone plate is written as
\begin{equation}
\label{eq:eq5}
U'_2(x_2)  = \sum_m \frac{V(x_2)}{\sqrt{|m|\pi}} e^{\frac{-i m k x_2^2}{2f}} U_2(x_2), 
\end{equation}
where $m$ = $\pm 1, \pm 3, ...$, and $f$ corresponds to the first order ($m$ = $+1$) focal length of the zone plate. Note that the zero-order term corresponding to the transmission of the incident radiation in the forward direction is not included here \cite{Howe07}. The $V(x_2)$, closely related to the size of the zone plate, stands for the correction factor of its amplitude transmission function \cite{Yash10}. With an ideal zone plate having an infinitely large area, $V(x_2)$ becomes a constant ($V(x_2)=1$ is assumed) and the associated spatial resolution becomes infinitesimal. 

Depending on the distance $d_3$, the complex amplitude of the X-ray wave field before the grating has two forms. If the grating is placed at the image plane of the source ($d_3=\frac{f(d_1+d_2)}{m(d_1+d_2)-f}$), the amplitude turns into an extreme value due to the use of the Dirac delta function at the beginning to simulate the point source. For the rest placement of the grating, the amplitude of the X-ray wave field is expressed by 
\begin{equation}
\label{eq:eq6}
\begin{aligned}
U_3(x_3) \approx & \sum_m \sqrt{\frac{f}{ i \lambda \pi \left| m\right| q}} e^{ik\left(d_1+d_2+d_3+\frac{1}{2q^2}S(x_3)\right)}, \\
S(x_3)   = & d_3 \left(m x_3\left(d_1+d_2\right)+f (\eta-x_3)\right)^2 \\ 
& + \left(d_1+d_2\right) \left(d_3 \eta m+f (x_3-\eta)\right)^2 \\
& - f m \left(d_3 \eta+x_3\left(d_1+d_2\right) \right)^2\\ 
& - 2 q^2 \phi \left(\frac{d_2 \eta \left(f-d_3 m\right)+d_3 f \eta+d_1 f x_3}{q}\right).
\end{aligned}
\end{equation}
where $q =  d_3\left(f-m \left(d_1+d_2\right)\right)+f\left(d_1+d_2\right)$. In particular, the aforementioned X-ray microscope equipped with a Talbot(-Lau) interferometer \cite{Take08, Taka17} works under this condition.

The amplitude of the wavefront after the grating is obtained by the product of $U_3(x_3)$ and the periodic transmission function of the grating $T(x_3)$. Considering a Ronchi grating with a duty cycle of 0.5, this equals to
\begin{equation}
\label{eq:eq7}
U'_3(x_3)  = T(x_3)U_3(x_3) = \sum_{n=-\infty}^{\infty} a_n e^{\frac{2 i n \pi x_3}{p}} U_3(x_3), 
\end{equation}
in which $T(x_3)$ is described by the form of Fourier series expansion, $p$ represents the grating period, and the coefficient $a_n$ is determined by the grating type and the diffraction order $n$.

The X-ray field reaching the detector plane is derived as
\begin{equation}
\label{eq:eq8}
\begin{aligned}
 U_4(x_4) \approx & \sum_{n=-\infty}^{\infty} \sum_m a_n \sqrt{\frac{f}{i \lambda \pi \left| m \right| \tau}} e^{ik\left(d_1+d_2+d_3+d_4+ Y(n, x_4) \right)},\\
 Y(n, x_4) = & - \frac{d_3}{2k p\tau} \left(4 \pi  \left(d_1+d_2\right) m n x_4-4 \pi  f n x_4+\eta^2 k m p\right) \\
& - \phi \left(\frac{\eta k p \left( \tau - d_1 \left(f-m(d_3+d_4)\right) \right)+d_1 f \left(k p x_4-2 \pi  d_4 n\right)}{k p \tau}\right) \\
& -\frac{2 \pi ^2 n^2 d_4}{k^2 p^2 \tau} \left(d_3 \left(f-\left(d_1+d_2\right) m\right)+\left(d_1+d_2\right) f\right) + \frac{f}{2\tau} (\eta-x_4)^2 \\
& - \frac{d_4 \eta}{2k p\tau} (\eta k m p-4 \pi  f n)- \frac{x_4}{2k p\tau}\left(d_1+d_2\right)  (k m p x_4-4 \pi  f n),
\end{aligned}
\end{equation}
with $\tau=\left( d_1+d_2 \right) \left(f-m(d_3 +d_4) \right)+\left(d_3+d_4\right) f$. Since $U_4(x_4)$ is generated from a point-shaped source, the influence of arbitrary shaped sources on the final amplitude can be estimated by the integration of $U_4(x_4)$.

Further, the order $m$ is set to +1 because the diffraction efficiencies of other odd orders decrease rapidly by a factor of $\frac{1}{m^2}$. Similarly, the diffraction order $n$ of the grating is limited to 0 and $\pm 1$ \cite{Yong20}. By the first-order approximation, the detected beam intensity, i.e., $\left| U_4(x_4) \right|^2$, is proportional to
\begin{equation}
\label{eq:eq9}
\begin{aligned}
I(x_4) \propto & \frac{1}{2} + \frac{4}{\pi ^2} +\frac{4}{\pi} \sin \left(\frac{\Theta_1 + \Phi_1-\Theta_2 - \Phi_2}{2}\right) \times \\
 & \cos \left(C x_4 + \frac{\left( \Theta_1 + \Phi_1+\Theta_2 + \Phi_2 +\pi\right)}{2} \right),
\end{aligned}
\end{equation}
\begin{equation*}
\begin{aligned}
\text{where} \ \ \ 
C & = \frac{2 \pi }{p M}\left(1- \frac{d_3}{f} + \frac{d_4}{d_1 M} \right) \\
\Theta_1 & = -\frac{\pi  d_4}{k p} C + \frac{2 \pi  d_4 \eta}{d_1 M p} \\ 
\Theta_2 & = \frac{\pi  d_4}{k p} C + \frac{2 \pi  d_4 \eta}{d_1 M p} \\
\Phi_1 & = k \left(\phi \left(-\frac{k p x_4-2 \pi  d_4}{k p M}\right)-\phi \left(-\frac{x_4}{M}\right)\right) \\ 
\Phi_2 & = k \left(\phi \left(-\frac{x_4}{M}\right)-\phi \left(-\frac{ k p x_4 + 2 \pi  d_4}{k p M }\right)\right). 
\end{aligned}
\end{equation*}

Introducing $d_s$ as the distance between the zone plate and the image plane of the source ($d_s < d_3$ in our case), the expression of the factor $C$ becomes 
\begin{equation}
\label{eq:eq10}
\begin{aligned}
C = \frac{2 \pi f \left(d_s-d_3\right)}{p M \left( d_2 \left(f-d_s\right)+f d_s \right)},
\end{aligned}
\end{equation}
whose value is highly sensitive to the location of the grating. When the magnification $M$ of the sample is sufficiently large, the object length $d_2$ approaches the focal length $f$, making the factor $C$ become $\frac{2 \pi \left(d_s-d_3\right)}{p M f}$. In addition, the period of the fringe pattern is equal to $2\pi/\left| C \right|$. And $\Theta_i$ ($i$=1,2) denotes the constant phase shift. Without the sample, the fringe visibility is maximized when the absolute value of the sine term in Eq.~(\ref{eq:eq9}) is equal to 1, which means that $d_4 C= kpw/2$, ($w=\pm 1, \pm 3, ...$).

To extract the phase information, the phase stepping (PS) approach is used. In practice, the grating is translated multiple times (denoted as an integer $M_{\text{ps}}$ $\geq$ 3) in the lateral direction by a certain distance $\Delta_{\text{ps}}$. This operation can be theoretically described by replacing $T(x_3)$ in Eq.~(\ref{eq:eq7}) with $T\left(x_3-\Delta_{\text{ps}}\right)$. After repeating the same procedure from Eq.~(\ref{eq:eq7}) to (\ref{eq:eq9}), the final intensity has a format of 
\begin{equation}
\label{eq:eq11}
\begin{aligned}
I^{(j)}_{\text{ps}}(x_4)  \propto & \frac{1}{2} + \frac{4}{\pi ^2} +\frac{4}{\pi} \sin \left(\frac{\Theta_1 + \Phi_1-\Theta_2 - \Phi_2}{2}\right) \times \\
 & \cos \left( \frac{2 \pi  j}{M_{\text{ps}}}+ C x_4 + \frac{\left( \Theta_1 + \Phi_1+\Theta_2 + \Phi_2 +\pi\right)}{2} \right), 
\end{aligned}
\end{equation}
where $\Delta_{\text{ps}} M_{\text{ps}}=p$ and $j=1,2,\cdots,M_{\text{ps}}$. Normally the detected intensity forms a sinusoidal curve called as the phase stepping curve, from which the phase shift can be calculated analytically,
\begin{equation}
\label{eq:eq12}
\begin{aligned}
\frac{1}{2}\left( \Phi_1 + \Phi_2 + \Delta(x_4) \right) = \tan^{-1}\left[ \frac{\sum_{j=1}^{M_{\text{ps}}} I^{(j)}\sin\left(\frac{2\pi j}{M_{\text{ps}}} \right)}{\sum_{j=1}^{M_{\text{ps}}} I^{(j)}\cos\left(\frac{2\pi j}{M_{\text{ps}}} \right)}\right] , 
\end{aligned}
\end{equation}
in which $\Delta(x_4)=2C x_4+\Theta_1 +\Theta_2 +\pi$ corresponds to the background measurement without the sample. Finally, the extracted phase signal is 
\begin{equation}
\label{eq:eq13}
\begin{aligned}
\frac{\Phi_1 + \Phi_2 }{2} = \frac{k}{2} \left( \phi \left(-\frac{x_4}{M}+\frac{\lambda  d_4}{ p M}\right) - \phi \left(-\frac{x_4}{M}-\frac{\lambda  d_4}{ p M}\right) \right), 
\end{aligned}
\end{equation}
which is formally consistent with the derivations as had been reported in Ref.~\cite{Yash09, Yash10}. It can be learned from this equation that the phenomenon of twin phase images would show up. These two enlarged (by a factor of $M$) phase images, splitted by a distance of $\frac{2\lambda  d_4}{ p M}$ from each other, have opposite amplitudes. And the magnitude of such positive and negative phase signals are reduced by $50\%$. If such separation becomes small enough compared with the resolution of the system $\epsilon$, such phase difference signals would be converted into the differential phase signal, namely,   
\begin{equation}
\label{eq:eq14}
\begin{aligned}
\frac{\Phi_1 + \Phi_2 }{2} \xlongrightarrow{\frac{\lambda  d_4}{ p M} < \epsilon} \frac{\lambda d_4}{p M} k \phi'(-\frac{x_4}{M}).
\end{aligned}
\end{equation}
Note that the coefficient $\frac{2\pi d_4}{p M}$ of $\phi'$ represents the sensitivity of the DPI signal. As the magnification $M$ increases, the sensitivity is significantly reduced. According to the Eq.~(\ref{eq:eq14}), there is always a competition between the separation and the sensitivity in realizing DPI since both of them are determined by the same factor of $\frac{\lambda d_4}{p M}$.

\section{Numerical simulation}

A series of 2D numerical simulations regarding the X-ray microscopic system shown in Fig.~\ref{fig:fig1} were conducted to verify the above theoretical derivations. Specifically, the Fresnel diffraction integral was calculated based on the discrete fast Fourier transform (DFFT) in Python. The transmission of X-rays through the optical components and the sample is imitated with the projection approximation. It is also assumed that the X-ray source is perfectly coherent and the detector has an ideal response without any crosstalk and photon shot noise. Besides, the resolution of the detector is determined by the discretization of the imaging field of view. To facilitate comparisons, the same imaging geometry and beam energy as listed in Ref.~\cite{Taka17} were taken into account. In all simulations, the X-ray energy was fixed at 8.04 keV, and the imaging system ran in the large field of view (LFOV) mode with a 10-fold sample magnification. The simulated imaging field of view is 120~$\mu$m$\times$120~$\mu$m with a zone plate having a resolution of 97.7~nm and a detector possessing a pixel dimension of 76.9~nm.

\begin{figure}[t]
\centering
\includegraphics[width=0.85\linewidth]{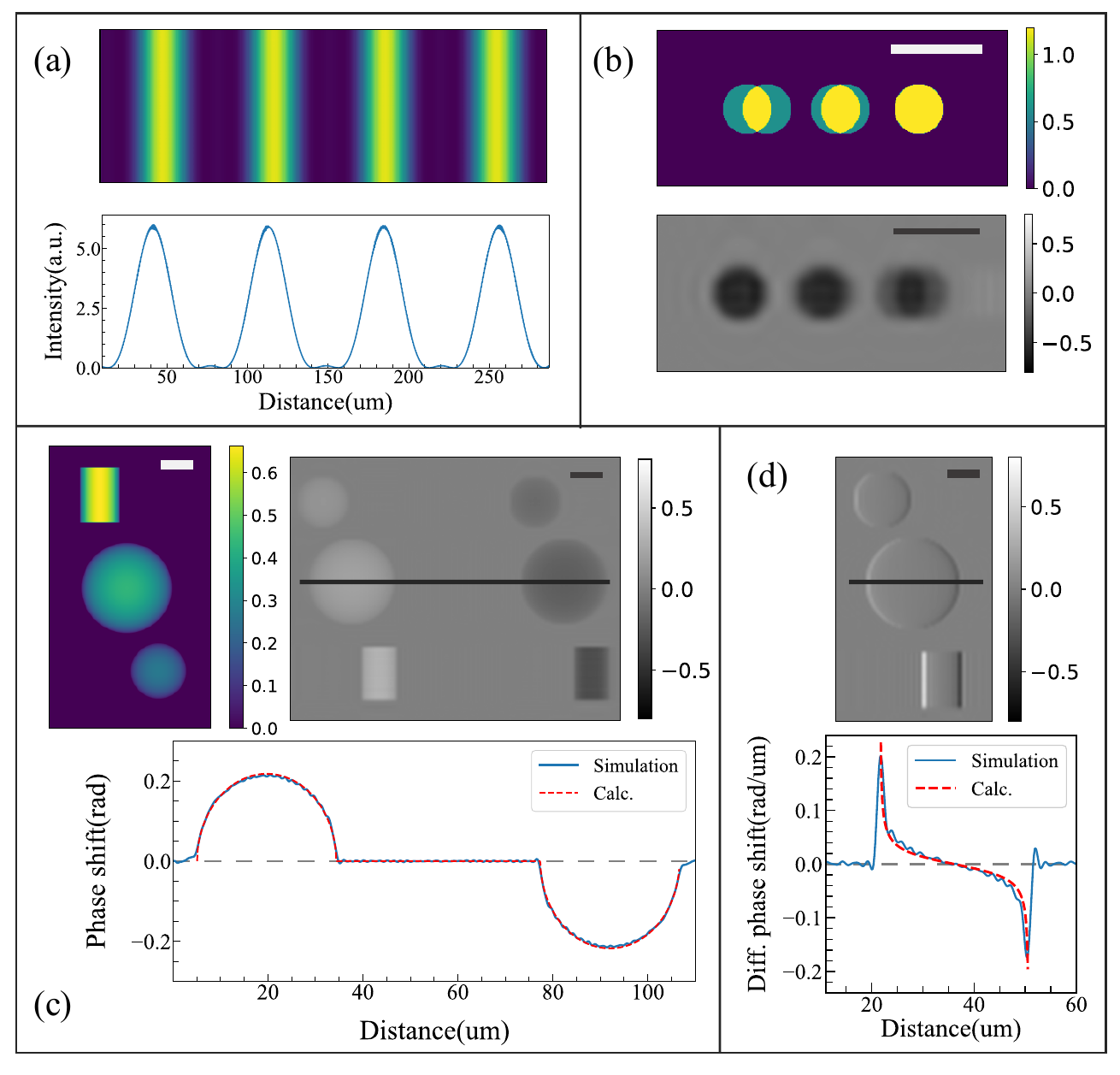}
\caption{Numerical simulation results: (a) the fringe distribution, (b) the phase difference imaging results to evaluate the spatial resolution, (c) the phase difference imaging results, and (d) the differential phase imaging results. The white and black scale bars denote 1~$\mu$m and 10~$\mu$m, respectively.}
\label{fig:fig2}
\end{figure}

The simulated fringe distribution without a sample is shown in Fig.~\ref{fig:fig2}(a). It has a period of 71.56~$\mu$m which can be obtained by Eq.~(\ref{eq:eq10}). This is consistent with the value reported in Ref.~\cite{Taka17}. The spatial resolution performance of the imaging system is evaluated in Fig.~\ref{fig:fig2}(b). The symmetrical structures in the figure are formed by overlapping two cylinders whose axes are parallel to the optical one. The offset distances associated with the three samples from left to right are 0.2~$\mu$m, 0.1~$\mu$m, and 0.0~$\mu$m, respectively. Due to the lens imaging procedure, an inverted image of the sample is found. The gray scale image shown here is actually the dark part of the phase difference image. Within it, the structure of 0.1~$\mu$m can hardly been seen since it is close to the resolution of the zone plate. In addition, the PDI results of different pure phase objects are displayed in Fig.~\ref{fig:fig2}(c). The sample contains two polystyrene spheres with diameters of 1.80~$\mu$m and 2.93~$\mu$m, and a rod of length 1.78~$\mu$m and diameter 1.27~$\mu$m. The corresponding maximum phase shifts for the Cu-K$\alpha$ X-ray photons are set to be 0.267~rad, 0.434~rad, and 0.662~rad, respectively. As shown in the bottom plot in Fig.~\ref{fig:fig2}(c), the measured separation between the twin phase images (72.32~$\mu$m) and the extracted phase shift from the simulation agree well with the theoretical calculations. The minor oscillation remained on the curve is most likely caused by the discontinuity of the signal when performing the DFFT. These tests also serve as a supplement to the previous theory, which indicates that including a finite-size zone plate would not affect the theoretical findings in this work. Finally, the DPI results of the same sample are presented in Fig.~\ref{fig:fig2}(d). The grating period considered in this simulation is equal to 100~$\mu$m. The obtained DPI signal is in good agreement with the theoretical estimates. The discrepancy along the edge of the sample is due to the fact that the coefficient $\frac{\lambda  d_4}{p M}$ cannot be infinitely small.

\begin{figure}[t]
\centering\includegraphics[width=0.9\linewidth]{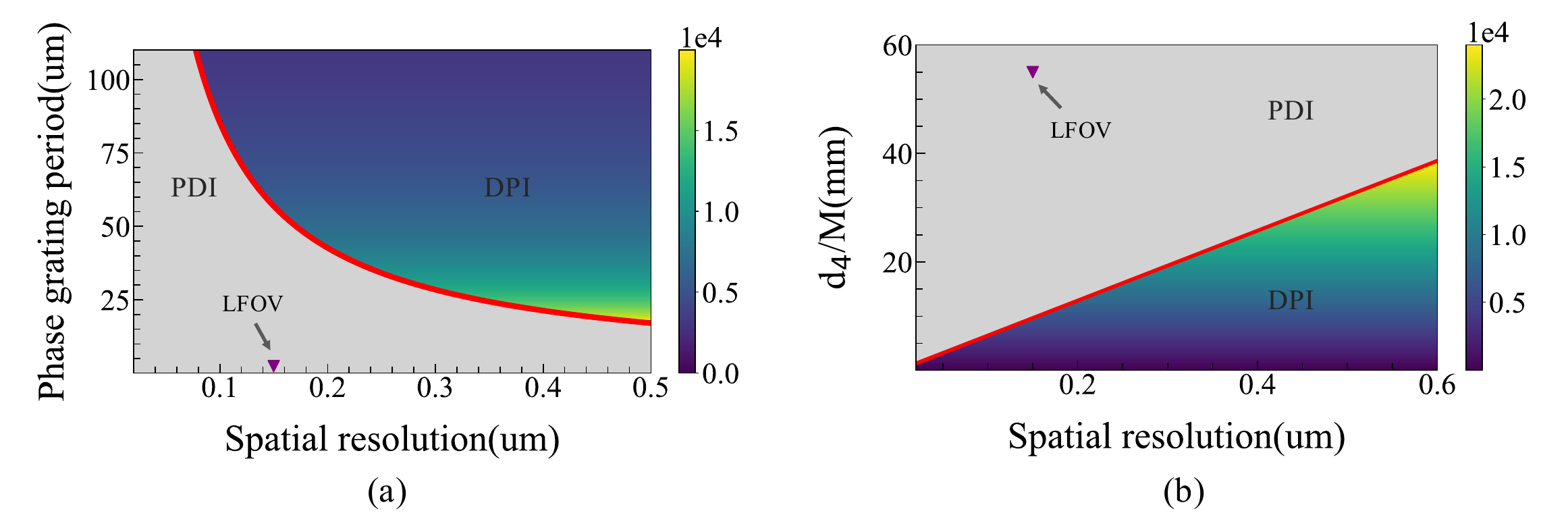}
\caption{Conversion conditions between PDI (gray area) and DPI (colored area) for the settings in Ref.~\cite{Taka17}: (a) with varied spatial resolutions and grating periods; (b) with varied spatial resolutions and the geometrical factor $d_4/M$. The colorbar denotes the sensitivity of the DPI. The inverted triangle corresponds to the LFOV mode.}
\label{fig:fig3}
\end{figure}

On the basis of Eq.~(\ref{eq:eq14}), the inherent unity and conversion of the PDI and DPI are further demonstrated in Fig.~\ref{fig:fig3}. It is clear that certain imaging conditions involving the beam condition, imaging geometry ($d_4/M$), and the grating period are required in order to realize the PDI or DPI, provided that the system spatial resolution is pre-defined. In addition, the figure shows that the sensitivity of DPI increases as the grating period gets reduced or the spatial resolution decreases. Whereas, the sensitivity of DPI is enhanced with a larger value of $\frac{d_4}{M}$. The red colored boundaries (solid lines) in the two subplots are obtained by assuming that $\frac{\lambda  d_4}{ p M} = \epsilon$. The boundaries would be shifted to the DPI side when $\frac{\lambda  d_4}{ p M} < \epsilon$.

\section{Discussion and Conclusion}

As demonstrated in the theoretical analyses, both the PDI and the DPI can be explained by the same imaging theory, upon which the PDI and DPI are unified. Meanwhile, certain phase signals, either PDI or DPI ones, can be retrieved under specific imaging conditions. By adjusting the imaging condition appropriately, the PDI can be converted to the DPI, or vise versa. For instance, increasing the grating period while maintaining other system settings would transform the X-ray microscope from the PDI mode to the DPI mode. However, our theoretical analysis also reveals that the sensitivity of DPI may be limited when pursuing a high image resolution. To this end, the parameters of an X-ray microscope system integrated with a grating interferometer need to be carefully optimized to meet the desired imaging applications.

In conclusion, we have developed a comprehensive theoretical framework based on the diffraction theory to explain the formation of the phase information in a grating-based X-ray microscope. Analyses demonstrate that the phase difference imaging and the differential phase imaging originate from a unified phase imaging theory and can be converted by changing certain conditions. Additionally, the impact of different optical components including the X-ray source can be analyzed rigorously by this imaging theory. In future, optimizations such as exploiting different shaped X-ray sources and varied grating types would be investigated for given X-ray microscope imaging tasks.

\begin{backmatter}

\bmsection{Funding} National Natural Science Foundation of China (12027812, 11804356), Youth Innovation Promotion Association of the Chinese Academy of Sciences (2021362). 

\medskip

\bmsection{Disclosures} The authors declare no conflicts of interest.

\medskip

\bmsection{Data availability} Data underlying the results presented in this paper are not publicly available at this time but may be obtained from the authors upon reasonable request.

\end{backmatter}

%%%%%%%%%%%%%%%%%%%%%%% References %%%%%%%%%%%%%%%%%%%%%%%%%

%%%%%%%%%% If using BibTeX:
\bibliography{sample}

\begin{thebibliography}{10}
\newcommand{\enquote}[1]{``#1''}

\bibitem{Momo03}
A.~Momose, S.~Kawamoto, I.~Koyama, Y.~Hamaishi, H.~Takai, and Y.~Suzuki,
  \enquote{Demonstration of {X}-ray {Talbot} interferometry,}
  {\protect\JournalTitle{Jpn. J. Appl. Phys., Part 2}} \textbf{42}, 866--8
  (2003).

\bibitem{Pfei06}
F.~Pfeiffer, T.~Weitkamp, O.~Bunk, and C.~David, \enquote{{Phase retrieval and
  differential phase-contrast imaging with low-brilliance X-ray sources},}
  {\protect\JournalTitle{Nat. Phys.}} \textbf{2}, 258--261 (2006).

\bibitem{Zhu10}
P.~Zhu, K.~Zhang, Z.~Wang, Y.~Liu, X.~Liu, Z.~Wu, S.~A. McDonald, F.~Marone,
  and M.~Stampanoni, \enquote{Low-dose, simple, and fast grating-based {X}-ray
  phase-contrast imaging,} {\protect\JournalTitle{Proceedings of the National
  Academy of Sciences}} \textbf{107}, 13576--13581 (2010).

\bibitem{Zhu14}
P.~Zhu, Z.~Zhu, Y.~Hong, K.~Zhang, W.~Huang, Q.~Yuan, X.~Zhao, Z.~Ju, Z.~Wu,
  Z.~Wei, S.~Wiebe, and L.~D. Chapman, \enquote{Common characteristics shared
  by different differential phase contrast imaging methods,}
  {\protect\JournalTitle{Appl. Opt.}} \textbf{53}, 861--867 (2014).

\bibitem{Momo20}
A.~Momose, H.~Takano, Y.~Wu, K.~Hashimoto, T.~Samoto, M.~Hoshino, Y.~Seki, and
  T.~Shinohara, \enquote{Recent progress in {X}-ray and neutron phase imaging
  with gratings,} {\protect\JournalTitle{Quantum Beam Science}} \textbf{4}
  (2020).

\bibitem{Xu20}
X.~Ji, R.~Zhang, K.~Li, and G.-H. Chen, \enquote{Dual energy differential phase
  contrast {CT} ({DE-DPC-CT}) imaging,} {\protect\JournalTitle{IEEE
  Transactions on Medical Imaging}} \textbf{39}, 3278--3289 (2020).

\bibitem{Zan21}
G.~Zan, S.~Gul, J.~Zhang, W.~Zhao, S.~Lewis, D.~J. Vine, Y.~Liu, P.~Pianetta,
  and W.~Yun, \enquote{High-resolution multicontrast tomography with an {X}-ray
  microarray anode{\textendash}structured target source,}
  {\protect\JournalTitle{Proceedings of the National Academy of Sciences}}
  \textbf{118} (2021).

\bibitem{Take08}
Y.~Takeda, W.~Yashiro, T.~Hattori, A.~Takeuchi, Y.~Suzuki, and A.~Momose,
  \enquote{Differential phase {X}-ray imaging microscopy with {X}-ray {Talbot}
  interferometer,} {\protect\JournalTitle{Applied Physics Express}} \textbf{1},
  117002 (2008).

\bibitem{Yash09}
W.~Yashiro, Y.~Takeda, A.~Takeuchi, Y.~Suzuki, and A.~Momose,
  \enquote{Hard-{X}-ray phase-difference microscopy using a {Fresnel} zone
  plate and a transmission grating,} {\protect\JournalTitle{Phys. Rev. Lett.}}
  \textbf{103}, 180801 (2009).

\bibitem{Yash10}
W.~Yashiro, S.~Harasse, A.~Takeuchi, Y.~Suzuki, and A.~Momose,
  \enquote{{Hard-{X}-ray phase-imaging microscopy using the self-imaging
  phenomenon of a transmission grating},} {\protect\JournalTitle{Phys. Rev. A}}
  \textbf{82}, 043822 (2010).

\bibitem{Kuwa11}
H.~Kuwabara, W.~Yashiro, S.~Harasse, H.~Mizutani, and A.~Momose,
  \enquote{Hard-{X}-ray phase-difference microscopy with a low-brilliance
  laboratory {X}-ray source,} {\protect\JournalTitle{Applied Physics Express}}
  \textbf{4}, 062502 (2011).

\bibitem{Beru12}
S.~Berujon, H.~Wang, I.~Pape, K.~Sawhney, S.~Rutishauser, and C.~David,
  \enquote{X-ray submicrometer phase contrast imaging with a {Fresnel} zone
  plate and a two dimensional grating interferometer,}
  {\protect\JournalTitle{Opt. Lett.}} \textbf{37}, 1622--1624 (2012).

\bibitem{Taka17}
H.~Takano, Y.~Wu, and A.~Momose, \enquote{{Development of full-field {X}-ray
  phase-tomographic microscope based on laboratory {X}-ray source},} in
  \emph{Developments in X-Ray Tomography XI,}  vol. 10391 B.~Müller and
  G.~Wang, eds., International Society for Optics and Photonics (SPIE, 2017),
  pp. 136 -- 144.

\bibitem{Taka18}
H.~Takano, Y.~Wu, J.~Irwin, S.~Maderych, M.~Leibowitz, A.~Tkachuk, A.~Kumar,
  B.~Hornberger, and A.~Momose, \enquote{Comparison of image properties in
  full-field phase {X}-ray microscopes based on grating interferometry and
  {Zernike's} phase contrast optics,} {\protect\JournalTitle{Applied Physics
  Letters}} \textbf{113}, 063105 (2018).

\bibitem{Taka19}
H.~Takano, K.~Hashimoto, Y.~Nagatani, J.~Irwin, L.~Omlor, A.~Kumar, A.~Tkachuk,
  Y.~Wu, and A.~Momose, \enquote{Improvement in quantitative phase mapping by a
  hard {X}-ray microscope equipped with a {Lau} interferometer,}
  {\protect\JournalTitle{Optica}} \textbf{6}, 1012--1015 (2019).

\bibitem{Wolf20}
A.~Wolf, M.~Schuster, V.~Ludwig, G.~Anton, and S.~Funk, \enquote{Maximum
  likelihood reconstruction for grating-based {X}-ray microscopy,}
  {\protect\JournalTitle{Opt. Express}} \textbf{28}, 13553--13568 (2020).

\bibitem{Howe07}
M.~Howells, C.~Jacobsen, T.~Warwick, and A.~Van~den Bos, \emph{Principles and
  Applications of Zone Plate X-Ray Microscopes} (Springer New York, New York,
  NY, 2007), pp. 835--926.

\bibitem{Yong20}
Y.~Ge, J.~Chen, P.~Zhu, J.~Yang, S.~Deng, W.~Shi, K.~Zhang, J.~Guo, H.~Zhang,
  H.~Zheng, and D.~Liang, \enquote{{Dual phase grating based {X}-ray
  differential phase contrast imaging with source grating: theory and
  validation},} {\protect\JournalTitle{Opt. Express}} \textbf{28}, 9786--9801
  (2020).

\end{thebibliography}

\end{document}